\documentclass[times,authoryear]{elsarticle}

\usepackage{jasr}
\usepackage{framed,multirow}

\usepackage{amssymb}
\usepackage{latexsym,booktabs}

\usepackage[switch]{lineno}
\usepackage{subcaption, graphicx}
\usepackage{url}
\usepackage{xcolor}
\definecolor{newcolor}{rgb}{.8,.349,.1}
\usepackage{mathtools}

\newcommand\Mycomb[2][^n]{\prescript{#1\mkern-0.5mu}{}C_{#2}}

\usepackage[citebordercolor=white]{hyperref}

\journal{Advances in Space Research}

\begin{document}

\verso{E. Thomas \textit{et. al}}

\begin{frontmatter}

\title{Wavelet analysis of possible association between sunspot number and rainfall over Kerala, India : A case study }

\author[1]{Elizabeth \snm{Thomas}}
\ead{shinuelz@yahoo.co.in}

\author[1]{S. \snm{Vineeth}}
\ead{vineethsmaikkattu2@gmail.com}

\author[1]{Noble P. \snm{Abraham}\corref{cor1}}
\ead{noblepa@gmail.com}
\cortext[cor1]{Corresponding author}

\address[1]{Department of Physics, Mar Thoma College, Kuttapuzha P. O. Tiruvalla, PIN 689103, Kerala, India}



\begin{abstract}
Global attention has been focused on extreme climatic changes.
This paper investigates the relationship between different phases of solar activity and extreme precipitation events in Kerala, India. Sunspot number and rainfall data were analysed over 122 years (1901-2022) and separated into winter, pre-monsoon, monsoon, and post-monsoon seasons on an annual scale. The study analysed climatic effects using 31-year mean values and conducted correlation and wavelet analyses (XWT and WTC). A negative correlation was observed in the winter and post-monsoon seasons, while positive correlations were seen in the pre-monsoon and monsoon seasons, all of which were statistically significant. Using cross-wavelet transform (XWT), the temporal relationship between sunspot number and rainfall values was investigated, revealing significant cross-power at an 8-12 year scale across all seasons. Wavelet coherence between the two data sets demonstrated significant correlation at the 2-4 and 4-8 year scales throughout the four seasons. Strong connections were evident at higher periods, such as the 8-16 year scale in the monsoon and post-monsoon seasons. The results show that the seasonal rainfall over Kerala is related to solar activity.

The solar phases of Solar Cycles 14-24 were determined for all seasons, and the years with excessive and insufficient rainfall were identified. It was observed that the descending phase had an impact on excess rainfall events during the winter and pre-monsoon seasons, while the ascending phase notably affected the monsoon and post-monsoon seasons. The study specifically examined the different magnetic polarities of sunspots in alternating solar cycles, focusing on even and odd cycles. It was found that extreme rainfall events were more frequent during the winter and pre-monsoon seasons in the even cycles, whereas in the odd cycles, they were more prevalent during the monsoon and post-monsoon seasons.  These findings are presented for the first time and may offer new perspectives on how different phases affect rainfall. This study suggests a physical link between solar activity and extreme precipitation in Kerala, which could increase predictability.

\end{abstract}


\end{frontmatter}


\section{Introduction}
\begin{figure*}[!ht]
\centering
\includegraphics[width=0.5\textwidth]{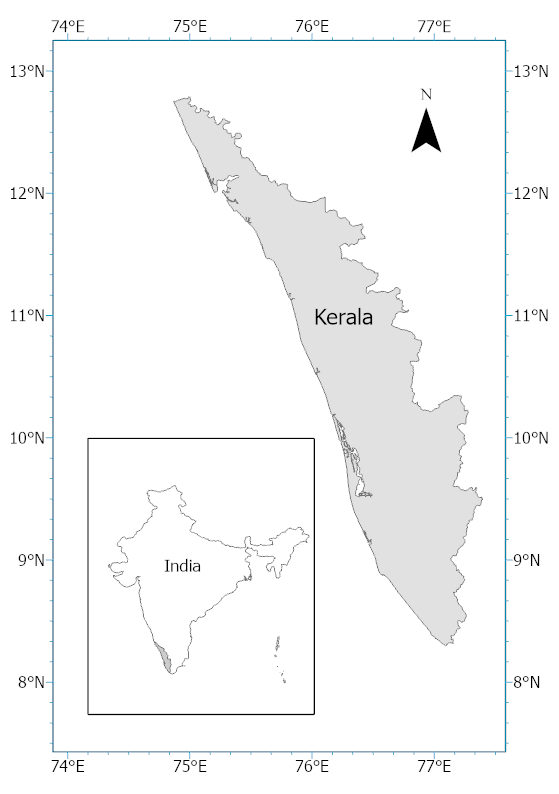}
\caption{Location map of Kerala}
\label{kerala}
\end{figure*}

Global climate change poses a hazard to human existence. The sun and anthropogenic factors exert a significant influence on weather and climate. The Sun's magnetic fields exhibit various spatial, temporal, and energetic phenomena. Sunspots, solar flares, solar wind, coronal mass ejections, etc., are all expressions of magnetic activity in the Sun, collectively known as solar activity \citep{Usoskin2017}. 
Sunspot number quantifies sunspots and is widely used because of its long-term availability. 
There has long been concern about how the sun affects precipitation on Earth. Precipitation in different parts of the world appears to be affected by the sun at various intervals. The effect of solar activity on rainfall varies with time scale and region, leading to both positive and negative correlations \citep{Tsiropoula2003, ZhaoJuanHanYan-BenandLi2004, Wasko2009, Mauas2011, Rampelotto2012}. Recently, few studies have been conducted on the relationship between solar and precipitation in China \citep{Zhai2017, YU2019,yan2022}, the United States \citep{Nitka2019}, Europe \citep{Laurenz2019}, Africa \citep{Mohamed2021}, Argentina \citep{HEREDIA2019105094}, Nepal \citep{tiwari2021}, and Northeast Asia \citep{yan2022}.

The economy, agriculture, and ecosystem in India could be seriously impacted by changing rainfall patterns \citep{DoranaluChandrashekar2017}. Many researchers have looked into the potential of a connection between solar activity and rainfall throughout India or in various regions \citep{jagannathan1973changes, Ananthakrishnan1984, Hiremath2004, Bhattacharyya2005, Agnihotri2011, Badruddin2015, Warrier2017, Thomas2022}. The direct and indirect effects were studied, and the results were often localised and contradicted other authors \citep{JAGANNATHAN1973,bhalme1981solar, Hiremath2006, Bhattacharyya2007, Lihua2007, Selvaraj2009,selvaraj2011study, Selvaraj2013,Hiremath2015,Malik2018, Thomas2023}.

Kerala is located at the southwest tip of India, bounded east by the Western Ghats and the west by the Arabian Sea. It extends between 8$^{\circ}$15$^{\prime}$ and 12$^{\circ}$50$^{\prime}$ north latitudes
and between 74$^{\circ}$50$^{\prime}$ and 77$^{\circ}$30$^{\prime}$ east longitudes. It shares boundaries with Karnataka in the
north, Tamil Nadu in the east, and the Arabian Sea in the west. Kerala
has a wet and tropical climate, and the major contribution is from the
southwest monsoon and post-monsoon. The diverse features of Kerala make it more susceptible to climate change. It is known as the "Gateway of summer monsoon". Studies of long-term rainfall variability revealed that rainfall during the southwest monsoon significantly reduced while rainfall during the post-monsoon rose \citep{Krishnakumar2009, Kothawale2017}. Recently, few studies have reported the influence of sunspot number on the rainfall over Kerala \citep{thomas2022impact, Thomas2022,Thomas2023}. 

It is crucial to evaluate how solar activity influences rainfall patterns in various parts of the world as this helps comprehend regional variations, enhancing our knowledge of climate change and its localised effects helping in extreme weather events forecasts. In Kerala, recent extreme rainfall events have resulted in landslides or floods that have claimed lives and destroyed property. In India, several studies have linked solar activity to extreme weather events (see, e.g. \cite{bhalme1981cyclic, Azad2011}). However, research on the influence of different solar phases over rainfall is limited. Therefore, looking at extreme rain over the Kerala region during different solar phases will be interesting.

 This paper studies the possible relation of rainfall over Kerala with sunspot number using Cross-wavelet transform (XWT) and Wavelet coherence (WTC). The solar phases of Solar Cycles SC14 - SC24 are identified, and their relation with extreme rainfall events is evaluated. 
Section \ref{data} discusses the data and methodology of analysis. Section \ref{results} presents the results and discussion about the wavelet analysis and occurrences of extreme rainfall events during different solar activity phases during different seasons. Section \ref{conclusions} presents the conclusions.

\section{Data and Methods}\label{data}

\subsection{Dataset} \label{dataset}

We have used 122 years (1901 – 2022) of data on rainfall in Kerala and sunspot numbers, covering eleven complete Solar cycles (SC14 – SC24). Sunspot number (SSN) is the commonly used solar index to measure solar activity and is taken from the World Data Center SILSO, Royal Observatory of Belgium, Brussels. The dataset is available at \url{http://www.sidc.be/silso/datafiles}. Rainfall (in mm) over Kerala (RF) is obtained from the India Meteorological Department (IMD) gridded rainfall data
(0.25$^{\circ}$ × 0.25$^{\circ}$) \citep{Pai2014}. The IMD data can be accessed from \url{https://www.imdpune.gov.in/lrfindex.php}. The location map of Kerala is shown in Figure \ref{kerala}.
India Meteorological Department (IMD) classifies the rainfall seasons of India as Winter (January-February), denoted JF, pre-monsoon (March-May), denoted MAM, monsoon (June-September), denoted JJAS and post-monsoon (October-December) denoted as OND \citep{Hiremath2004, Hiremath2006, Bankoti2011}. This study divides the rainfall values into four seasons: JF, MAM, JJAS, and OND. The sunspot number values for each season are averaged and used throughout the study. 

\subsection{Methodology of analysis} \label{methodology}

Spearman correlation is a statistical method used to find the strength between the ranks of the raw values and how strong the relation is. The correlation coefficient values vary between -1 to 1, and its significance can be calculated. It is more robust than the linear Pearson method and is widely used \citep{hiremath2004influence,hiremath2006influence,Bankoti2011}.
Wavelet analysis is a powerful tool for analysing localised power variations within a time series, breaking the time series into a time-frequency space. Morlet’s wavelet is often preferred for feature extraction due to its balanced localisation of time and frequency \citep{grinsted, Torrence1998}. In this work, we perform Cross-wavelet analysis (XWT) and wavelet transform coherence (WTC). Cross-wavelet transform (XWT) is applied to the time series of sunspot number and rainfall data to reveal areas of common power and relative phases between them. Wavelet coherence (WTC) is performed to identify the localised correlation between the two time series in time-frequency space. For more details and specific formulas, refer \citep{grinsted}.
To study the influence of solar phases on the rainfall over Kerala, sunspot data of each season is considered, and the solar phases corresponding to Solar Cycles (SC14 - SC24) are classified using the technique outlined in \citep{SAWADOGO2023}. According to it, the maximum sunspot number ${S{N}_{max}}$ of each solar cycle is first calculated. The different phases are determined as follows: (i) minimum phase: $SN(t) < 0.122 \times S{N}_{max}$ (ii) increasing phase: $0.122 \times
S{N}_{max} \le SN(t)$ (iii) maximum phase: $SN(t) > 0.73 \times S{N}_{max}$ and (iv) decreasing phase: $0.73 \times S{N}_{max} \ge SN(t) > S{N}_{min}$ (next cycle). 

The years of excess and deficit rainfall over Kerala are identified to study the relationship between extreme rainfall events and solar activity. For that, the mean ($\mu$) and standard deviation ($\sigma$) of rainfall RF during all the seasons (JF, MAM, JJAS, and OND) are determined. A year i is labelled as extreme rainfall year when $R_{i} \geq (\mu + \sigma)$ and a year labelled as deficit rainfall year when 
$R_{i} \leq (\mu + \sigma)$, where $R_{i}$ is the rainfall of that year, $ i,k\in\mathbb{R} $ \citep{Azad2011}. In this study, k is defined as one.

\section{Results and discussions} \label{results}
\subsection{Variation of Sunspot number (SSN) and rainfall (RF)}
\begin{figure*}[!ht]
\centering
  \begin{subfigure}{0.4\textwidth}
    \includegraphics[width=\linewidth]{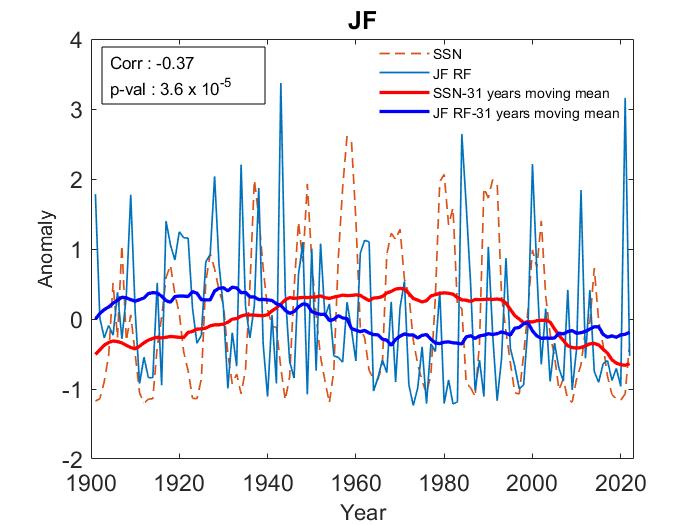}
    \caption{} \label{fig:2a}
  \end{subfigure}%
  \begin{subfigure}{0.4\textwidth}
    \includegraphics[width=\linewidth]{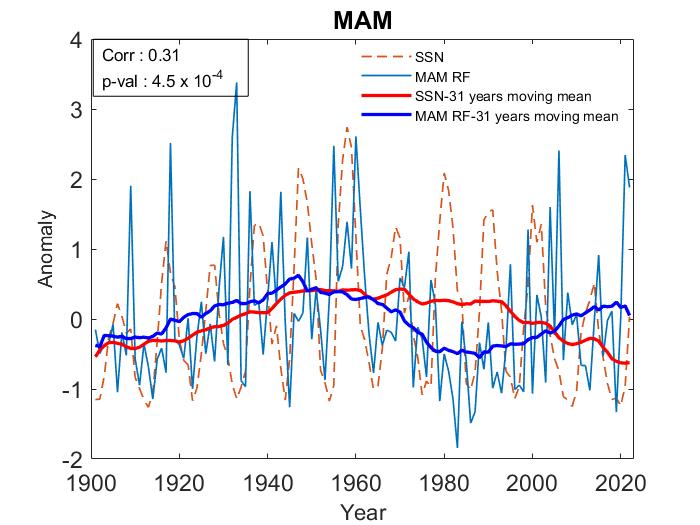}
    \caption{} \label{fig:2b}
  \end{subfigure}  \\
  \begin{subfigure}{0.4\textwidth}
    \includegraphics[width=\linewidth]{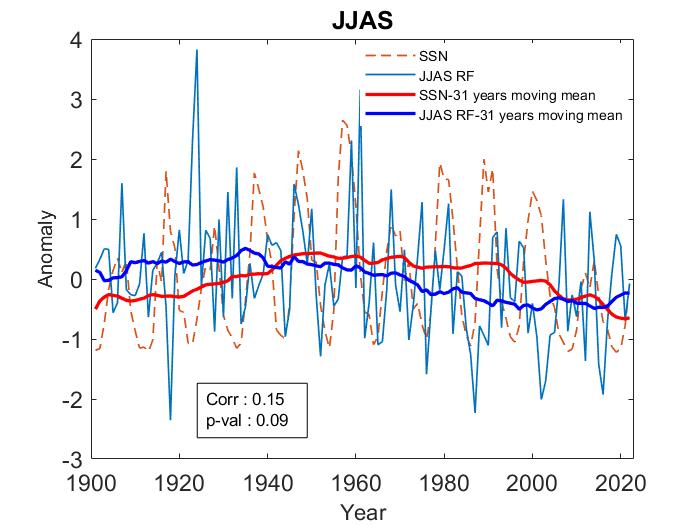}
    \caption{} \label{fig:2c}
  \end{subfigure}%
  \begin{subfigure}{0.4\textwidth}
    \includegraphics[width=\linewidth]{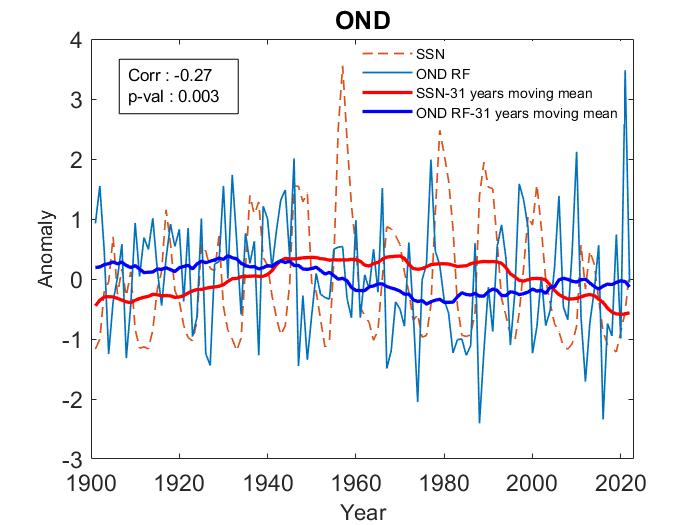}
    \caption{} \label{fig:2d}
  \end{subfigure}  \\
\caption{Standardized anomaly of sunspot number (SSN) and rainfall (RF) and their 31-years mean values during (a) JF (b) MAM (c) JJAS and (d) OND seasons. The correlation and p-value is calculated for the 31-years moving mean.}
\label{timeseries31mean}
\end{figure*}

The sunspot number and rainfall values corresponding to all seasons are standardised to easily compare results \citep{Chaudhuri2015, Thomas2023}. The standardised anomaly of rainfall (RF) and sunspot number (SSN) during JF, MAM, JJAS, and OND seasons is presented in Figure. \ref{timeseries31mean}. To understand the climate change effects, roughly 30 years average is considered \citep{Barde2023}. So, the 31-year moving average of sunspot number and rainfall data during all the seasons are calculated, and the Spearman correlation of these mean values is performed. In this method of 31 years moving average, 11-year and 22-year periods of the Sun are filtered out. Spearman correlation between the 31 years moving averaged rainfall values and sunspot are calculated along with its significance (Figure. \ref{timeseries31mean}). JF and OND seasons show a significant negative correlation, i.e., -0.37 and -0.27 respectively. JJAS showed a weak correlation, i.e., 0.15, compared to all seasons. A significant positive correlation is observed during the MAM season, i.e., 0.31.
\cite{Hiremath2004} reported a positive significant correlation between Indian rainfall variability and sunspot number during spring and southwest monsoon seasons. While considering cycle-to-cycle variations, another study showed that high solar activity would lower Indian rainfall variability and vice-versa. \citep{Hiremath2006} .\cite{Bankoti2011} performed correlative studies of two, four, and six-point moving averages of all India homogeneous rainfall with different solar activity features (sunspot number, solar active prominences, and Ha solar flares). Both positive and negative correlation results were observed for the seasonal and annual data. When the correlation values (positive or negative) were high, the significance was low and vice-versa.

\subsection{Cross-wavelet transform (XWT) and Wavelet coherence (WTC)}
Cross-wavelet transform is carried out between the 31 years averaged values of sunspot number and rainfall to statistically estimate the covariance level between them. Regions of common power are given in black contours with the phase relationship. 

\begin{figure*}[!ht]
\centering
  \begin{subfigure}{0.4\textwidth}
    \includegraphics[width=\linewidth]{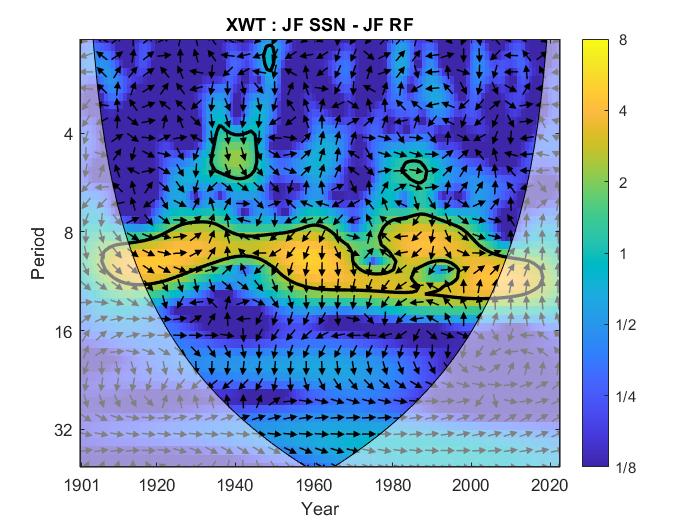}
    \caption{} \label{fig:2a}
  \end{subfigure}%
  \begin{subfigure}{0.4\textwidth}
    \includegraphics[width=\linewidth]{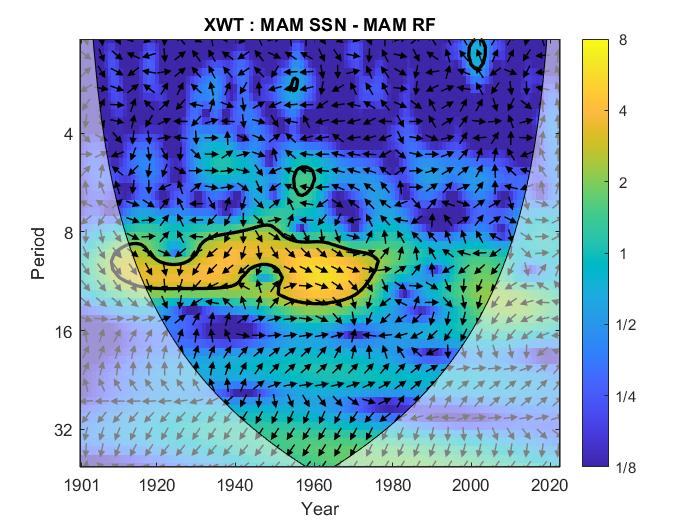}
    \caption{} \label{fig:2b}
  \end{subfigure}  \\
  \begin{subfigure}{0.4\textwidth}
    \includegraphics[width=\linewidth]{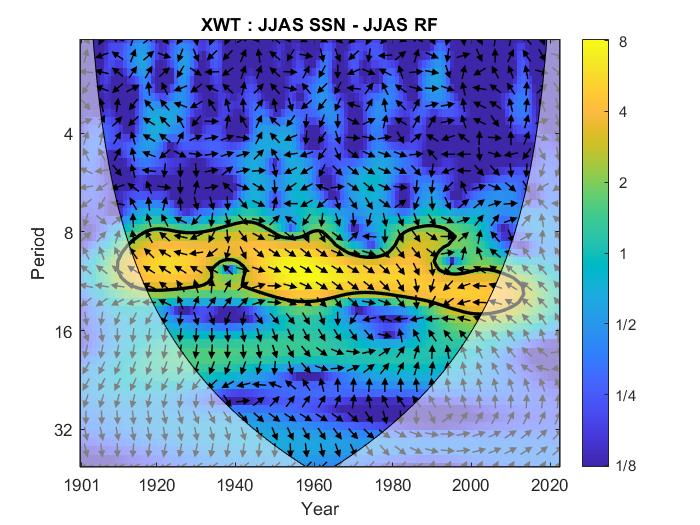}
    \caption{} \label{fig:2c}
  \end{subfigure}%
  \begin{subfigure}{0.4\textwidth}
    \includegraphics[width=\linewidth]{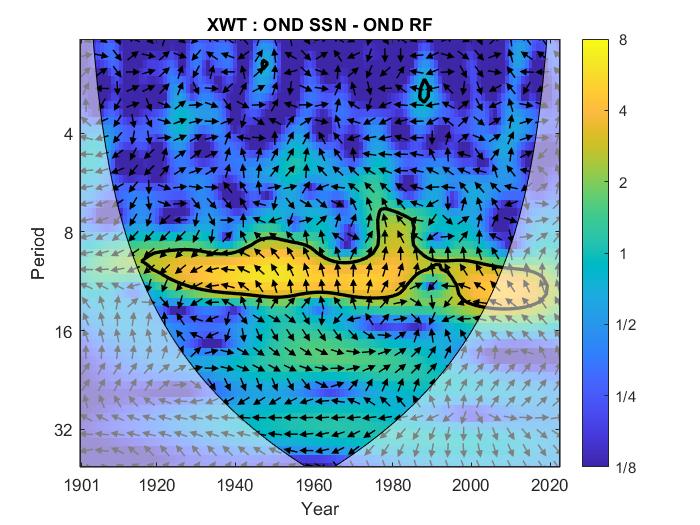}
    \caption{} \label{fig:2d}
  \end{subfigure}  \\
\caption{Wavelet cross-spectra between sunspot number (SSN) and rainfall (RF) corresponding to seasonal months (a) JF, (b) MAM, (c) JJAS, and (d) OND. The thick black contour
indicates the 5\% significance level and the cone of influence (COI) is shown as a black line.}
\label{xwt}
\end{figure*}

Figure \ref{xwt} illustrates the wavelet cross-spectra representing the relationship between sunspot number and rainfall values across all seasons. A bold black contour denotes a 5\% significance level, and the arrows depict the relative phase relationship between the two variables. Right-pointing arrows indicate an in-phase relationship, while left-pointing arrows signify an anti-phase relationship.

From the above figure, significant cross-power is observed at the 8–16 years scale for all seasons, suggesting a possible connection between sunspot numbers and rainfall and different seasons exhibited different patterns of areas of common power. The MAM season showed less cross-power area than the other seasons. A similar result is reported in another study over Kerala using different datasets of rainfall values \citep{Thomas2022,Thomas2023}. However, no definitive information regarding the phase could be inferred from the arrows, as phase reversals were noticeable at different times. Similar findings were documented by \cite{SouzaEcher2008} in Pelotas and \cite{Rampelotto2012} in Santa Maria, southern Brazil when investigating the variability of rainfall and its potential association with solar activity. These studies reported significant cross-power during the 11-year solar cycle period. Additionally, \cite{Nazari-Sharabian2020} identified similar cross-power results in their analysis of different stations in Iran. Recently, studies relating solar activity and monthly rainfall in Pokhara and Kathmandu cities in Nepal revealed similar cross-wavelet results \citep{Gautam2024}.

\begin{figure*}[t]
\centering
  \begin{subfigure}{0.4\textwidth}
    \includegraphics[width=\linewidth]{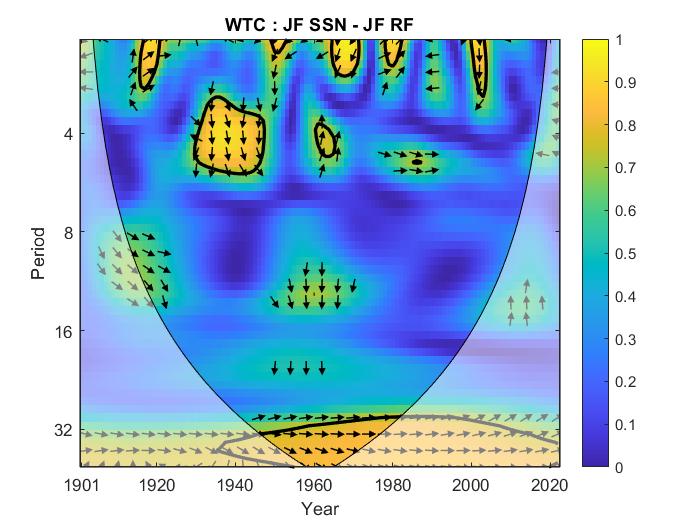}
    \caption{} \label{fig:2a}
  \end{subfigure}%
  \begin{subfigure}{0.4\textwidth}
    \includegraphics[width=\linewidth]{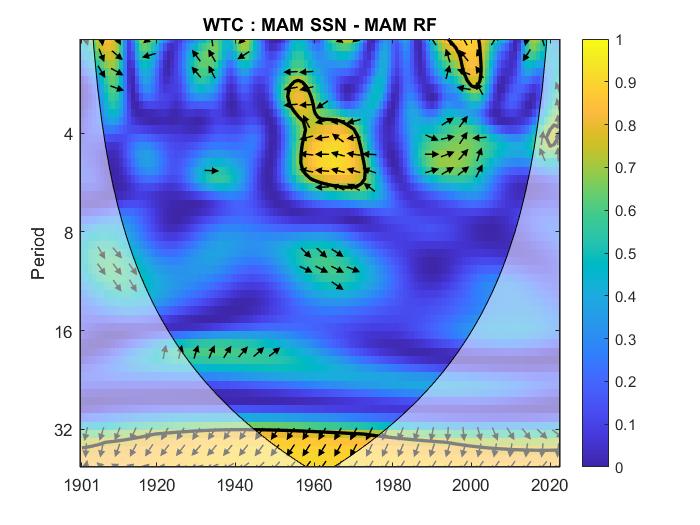}
    \caption{} \label{fig:2b}
  \end{subfigure}  \\
  \begin{subfigure}{0.4\textwidth}
    \includegraphics[width=\linewidth]{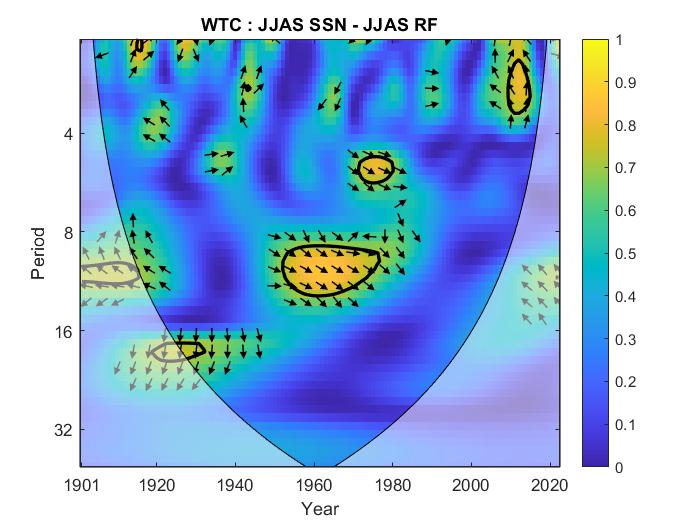}
    \caption{} \label{fig:2c}
  \end{subfigure}%
  \begin{subfigure}{0.4\textwidth}
    \includegraphics[width=\linewidth]{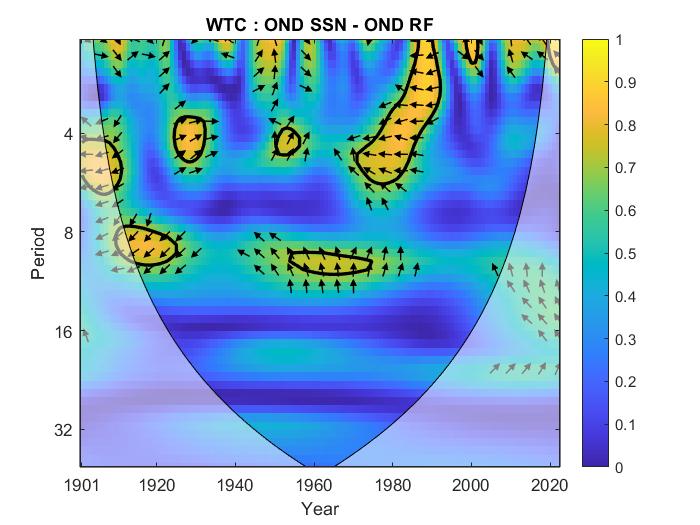}
    \caption{} \label{fig:2d}
  \end{subfigure}  \\
\caption{Wavelet coherence spectra between sunspot number (SSN) and rainfall (RF) corresponding to (a) JF (b) MAM (c) JJAS and (d) OND seasons. The thick black contour indicates the 5\% significance level and the cone of
influence (COI) is shown as a black line.}
\label{wtc}
\end{figure*}

Wavelet coherence analysis is conducted to statistically determine the connection between 31 years of averaged sunspot numbers and rainfall over Kerala during all the seasons. The dark contours in the graph indicate significant periods of coherence of a robust red noise process at the 5\% significance level \citep{grinsted}. The results of the wavelet coherence spectra throughout all seasons are presented in Figure. \ref{wtc}.

Significant high coherence is observed with intermittent power at 2-4 and 4-8 years scale during all four seasons. Significant coherence is visible at higher periods, i.e., at 8-16 years scale during JJAS and OND seasons, and above 32 years scale during JF and MAM seasons. Relevant phase information could not be obtained from the figure. Similar coherence studies have been carried out using sunspots and rainfall over Kerala using seasonal and annual values \citep{Thomas2022,Thomas2023} and recently in Nepal too \citep{Gautam2024}.

\subsection{Solar phase studies}
\begin{figure*}[t]
\centering
\includegraphics[width=0.8\linewidth]{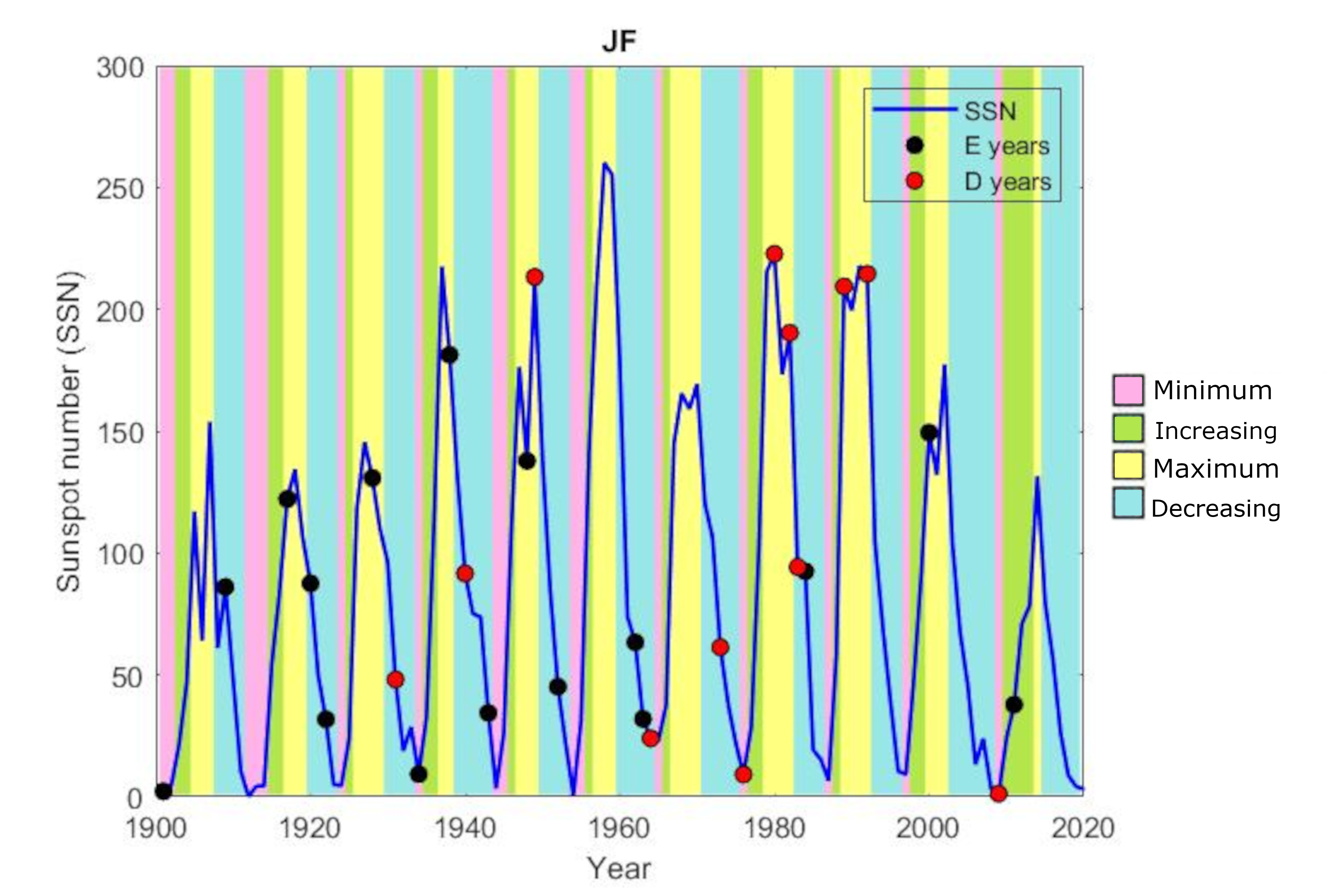}
\caption{Solar phases during Solar cycles 11-24 along with extreme years of rainfall, during the JF season. E denotes excess rainfall and D denotes deficit rainfall.}
\label{extremejf}
\end{figure*}
\begin{figure*}[t]
\centering
\includegraphics[width=0.8\linewidth]{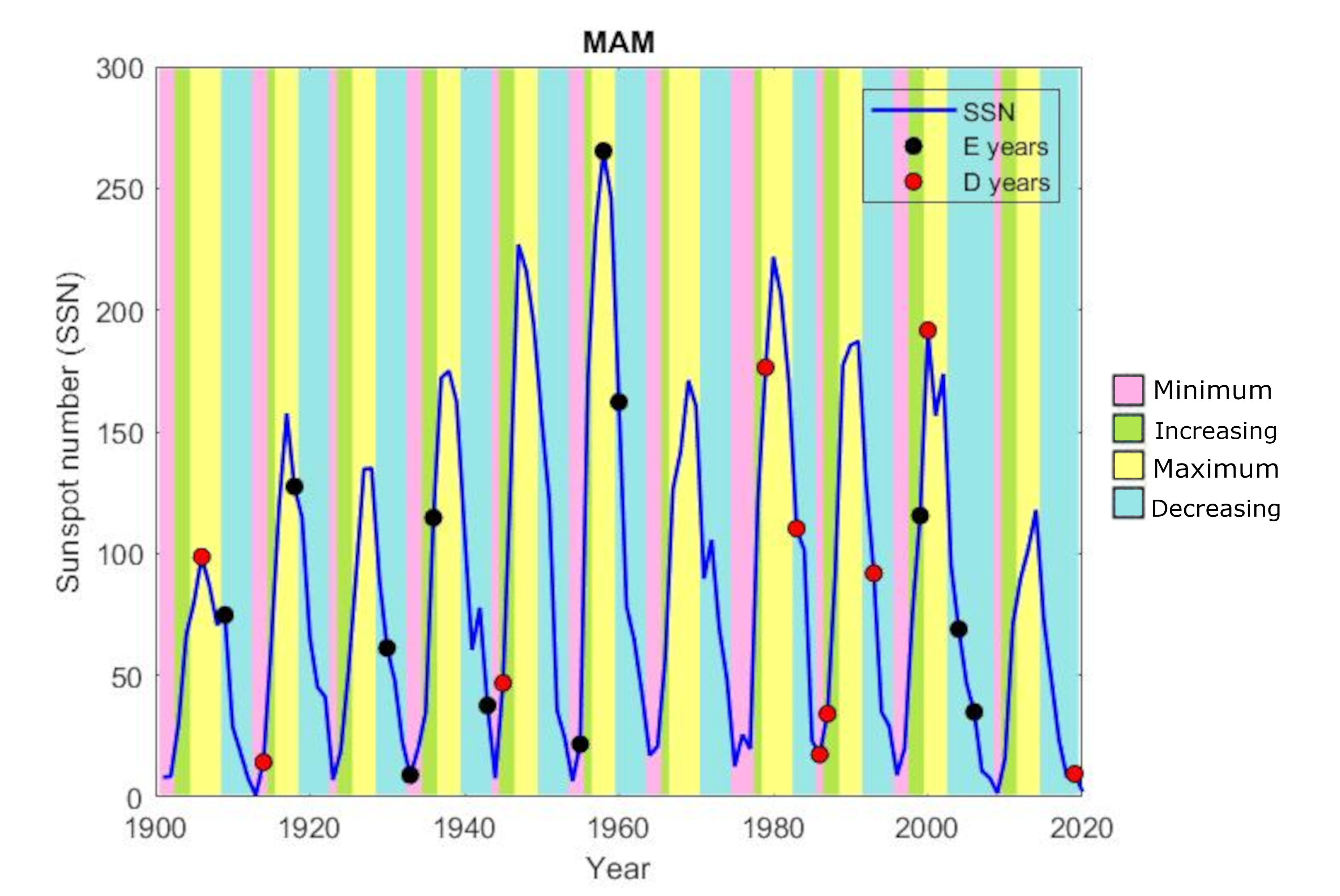}
\caption{Solar phases during Solar cycles 11-24 along with extreme years of rainfall, during the MAM season. E denotes excess rainfall and D denotes deficit rainfall.}
\label{extrememam}
\end{figure*}
\begin{figure*}[t]
\centering
\includegraphics[width=0.8\linewidth]{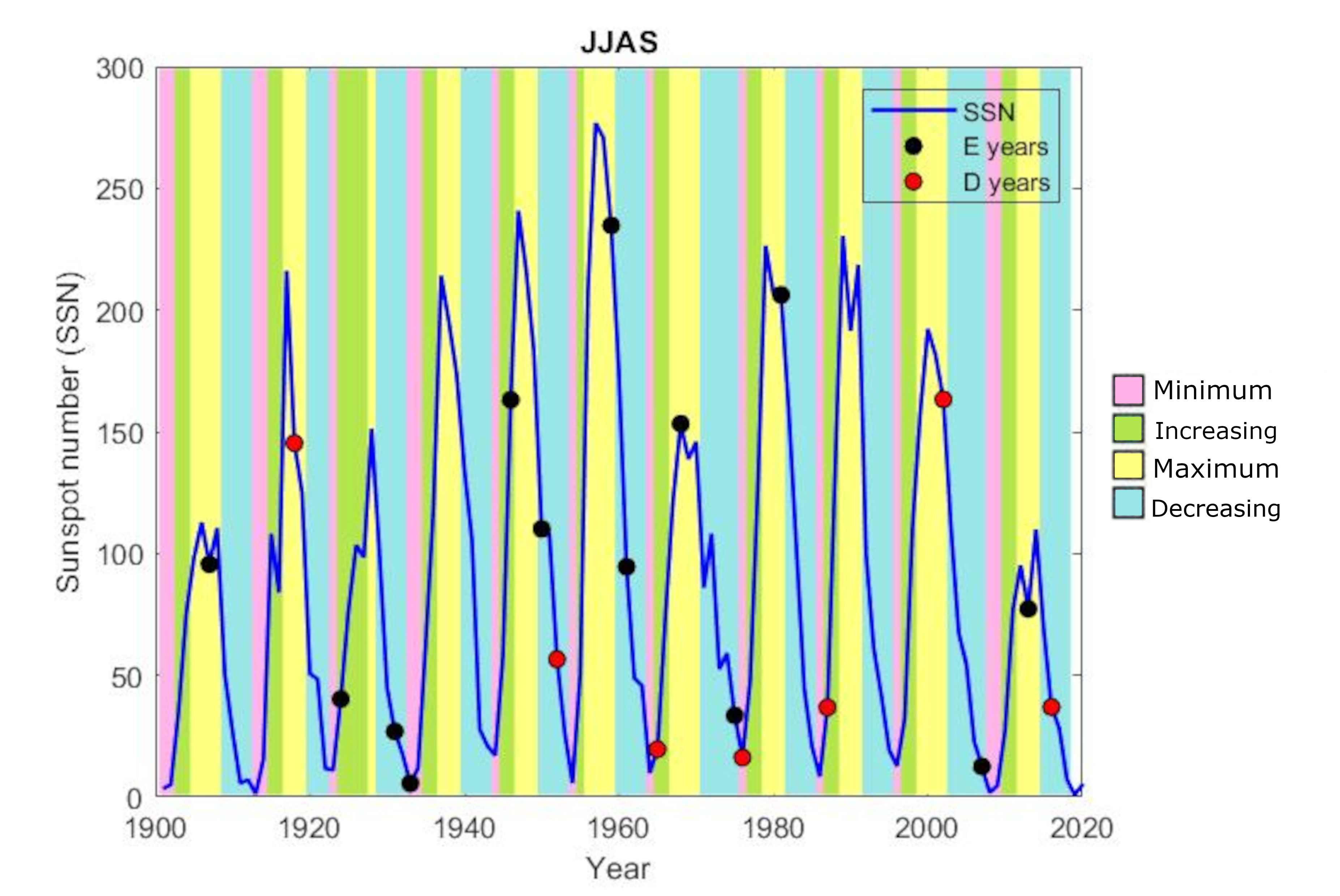}
\caption{Solar phases during Solar cycles 11-24 along with extreme years of rainfall, during the JJAS season. E denotes excess rainfall and D denotes deficit rainfall.}
\label{extremejjas}
\end{figure*}
\begin{figure*}[t]
\centering
\includegraphics[width=0.8\linewidth]{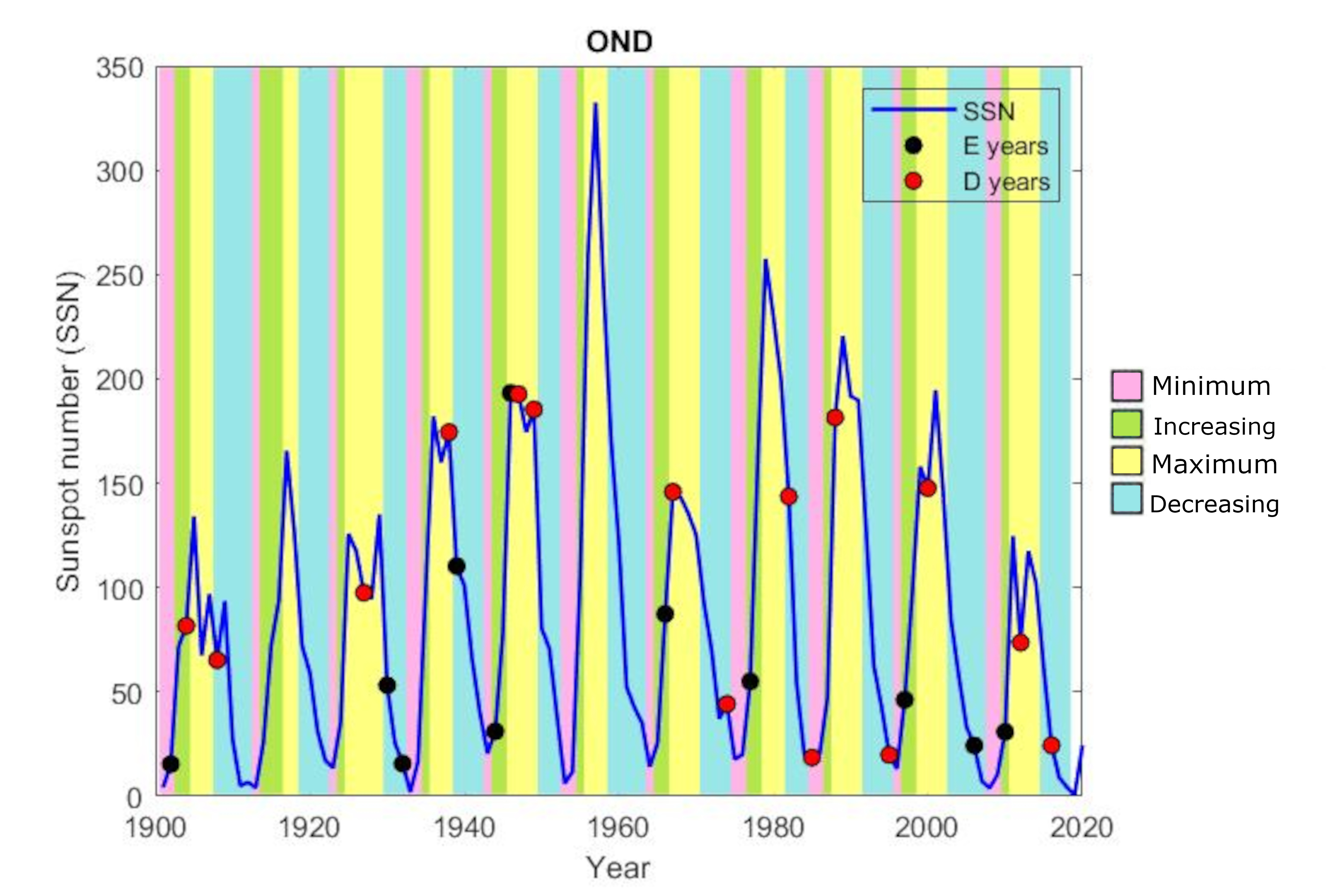}
\caption{Solar phases during Solar cycles 11-24 along with extreme years of rainfall, during the OND season. E denotes excess rainfall and D denotes deficit rainfall.}
\label{extremeond}
\end{figure*}

Attempts are made to understand the possible relationship between solar phases and extreme precipitation events in Kerala during different seasons (JF, MAM, JJAS, and OND). For that, the years of excess and deficit rainfall are first identified, as explained in Section \ref{data} \citep{Azad2011}. 
The solar phases of Solar Cycles 14-24 are determined using the criterion given in Section \ref{data} from the sunspot data for all seasons. Tables \ref{phasejf}, \ref{phasemam}, \ref{phasejjas}, and \ref{phaseond} show the classification of phases during the JF, MAM, JJAS, and OND seasons. 

The relative timing of solar activity and extreme rainfall are evaluated, corresponding to different seasons. Figure \ref{extremejf}, \ref{extrememam}, \ref{extremejjas}, and \ref{extremeond} shows the extreme rainfall events during JF, MAM, JJAS, and OND seasons, along with the phase classifications. The black circle represents excess rainfall years, and the red circle represents deficient rainfall years. The decreasing and minimum phases are combined as the descending phase, and the increasing and maximum phases as the ascending phase. 

The present study covers 11 complete solar cycles (Solar cycles 14-24), starting from 1901. On numbering these cycles sequentially from 1 to 11, our study covers six odd and five even cycles. Extreme rainfall events during odd and even cycles are also evaluated. 

\subsection{JF season}
A total of 16 excess and 12 deficit rainfall years are noted during the JF season and are listed below :
\begin{itemize}
\item Excess (E) years: 1901, 1909, 1917, 1920, 1922, 1928, 1934, 1938, 1943, 1948, 1952, 1962, 1963, 1984, 2000, 2011.
\item Deficit (D) years: 1931, 1940, 1949, 1964, 1973, 1976, 1980, 1982, 1983, 1989, 1992, 2009
\end{itemize}
 
Figure \ref{extremejf} denotes the extreme rainfall years during the JF season plotted on the SSN curve. The different phases during the solar cycles 14-24 are also indicated. During this season, excess rainfall events are observed more than deficit ones. Out of 16 excess rainfall years observed, 10 occurred during the descending phase and 6 during the ascending phase of the solar cycles considered. In the case of 12 deficit rainfall years, 7 occurred during the descending phase and 5 during the ascending phases of the solar cycles. The significance of these results is roughly evaluated using a binomial distribution \citep{Ananthakrishnan1984}. The hypothesis is taken such that the excess or deficit years of rainfall occur randomly without connection with the sunspot cycle. Considering all eleven solar cycles, the mean duration of a solar cycle was calculated as 10 years, the duration of the ascending phase as 5 years, and that of the descending phase as 5 years. Therefore, for random events, the probability of occurrence during the ascending and descending phases are 5/10 each. The probability of 10 or less out of 16 excess years occurring in the descending phase is 
\begin{equation}
    \sum_{r=0}^{10} \Mycomb[16]{r} \left( \frac{5}{10}\right)^r \left( \frac{5}{10}\right)^{16-r} \approx 0.895
\end{equation}  

The probability of occurrence of 10 years of excess rainfall during the descending phase is around 0.895, and occurrence of 7 years of deficit rainfall during the descending phase is around 0.806, which is not unusual. 

While studying the even and old cycles, it was observed that even cycles tend to have more extreme rainfall occurrences (excess/deficit) than the old cycles. Another observation from Figure \ref{extremejf} is that 8 out of 12 deficit years are over when the solar activity is at minimum or maximum peaks for the solar cycles.

\begin{table*}[!ht]
   \centering
  \caption{Solar cycles 14-24 classification of phases, during JF season}
     \label{phasejf}
    \begin{tabular}{ccccc} \toprule
       Solar cycle & Minimum & Maximum & Increasing & Decreasing \\ \midrule
       14 & 1901-1902 & 1905-1907 & 1903-1904 & 1908-1911 \\
       15 & 1912-1914 & 1917-1919 & 1915-1916 & 1920-1923 \\
       16 & 1924 & 1926-1929 & 1925 & 1930-1933 \\
       17 & 1934 & 1937-1938 & 1935-1936 & 1939-1943 \\
       18 & 1944-1945 & 1947-1949 & 1946 & 1950-1953 \\
       19 & 1954-1955 & 1957-1959 & 1956 & 1960-1964 \\
    20 & 1965 & 1967-1970 & 1966 & 1971-1975\\
       21 & 1976 & 1979-1982 & 1977-1978 & 1983-1986 \\
      22 & 1987 & 1989-1992 & 1988 & 1993-1996  \\
       23 & 1997 & 2000-2002 & 1998-1999 & 2003-2008   \\
       24 & 2009 & 2014 & 2010-2013 & 2015-2019 \\ \bottomrule     
  \end{tabular}
   \end{table*}

\subsection{MAM season}

During the MAM season, 12 excess and 10 deficit rainfall years are observed and listed below :
\begin{itemize}
\item Excess (E) years: 1909, 1989, 1930, 1933, 1936, 1943, 1955, 1958, 1960, 1999, 2004, 2006
\item Deficit (D) years: 1906, 1914, 1945, 1979, 1983, 1986, 1987, 1993, 2000, 2019
\end{itemize}

The excess and deficient rainfall years during the MAM season are indicated on the SSN curve, with different solar phases in Figure \ref{extrememam}. Out of the 12 excess rainfall years observed in this season, 8 occurred during the decreasing phase and 4 during the ascending phase of different solar cycles. However, the deficit years of rainfall were visible equally in both phases. The probability of occurrence of excess rain during the descending phase is around 0.927, which is not unusual. It was observed that even cycles have more extreme rainfall occurrences (excess/deficit) than old cycles, similar to the JF season.

\begin{table*}[!ht]
   \centering
  \caption{Solar cycles 14-24 classification of phases, during MAM season}
     \label{phasemam}
    \begin{tabular}{ccccc} \toprule
      Solar cycle & Minimum & Maximum & Increasing & Decreasing \\ \midrule
       14 & 1901-1902 & 1905-1908 & 1903-1904 & 1909-1912 \\
       15 & 1913-1914 & 1916-1918 & 1915 & 1919-1922 \\
       16 & 1923 & 1926-1928 & 1924-1925 & 1929-1932 \\
       17 & 1933-1934 & 1937-1939 & 1935-1936 & 1940-1943 \\
       18 & 1944 & 1947-1948 & 1945-1946 & 1950-1953 \\
       19 & 1954-1955 & 1957-1959 & 1956 & 1960-1963 \\
    20 & 1964-1965 & 1967-1970 & 1966 & 1971-1974\\
       21 & 1975-1977 & 1979-1982 & 1978 & 1983-1985 \\
      22 & 1986 & 1989-1991 & 1987-1988 & 1992-1995  \\
       23 & 1996-1997 & 2000-2002 & 1998-1999 & 2003-2008   \\
       24 & 2009 & 2012-2014 & 2010-2011 & 2015-2019 \\ \bottomrule      
  \end{tabular}
   \end{table*}
   
   \subsection{JJAS season}
   In the case of the JJAS season, 13 excess and 7 deficit rainfall events are seen and they are listed below.
 \begin{itemize}
\item Excess (E) years: 1907, 1924, 1931, 1933, 1946, 1950, 1959, 1961, 1968, 1975, 1981, 2007, 2013
\item Deficit (D) years: 1918, 1952, 1965, 1976, 1987, 2002, 2016
\end{itemize}   
    
A plot of the extreme rainfall years during the JJAS season is shown in Figure \ref{extremejjas}. There are also indications of the phases during solar cycles 14-24. During this season, 13 excess rainfall years have been recorded, 6 in the decreasing phase and 7 in the ascending phase. Out of the total 7 deficit years, 4 were visible during the ascending phase and 3 during the descending phase. Excess rainfall during the ascending phase has an occurrence probability of around 0.709, while deficit rainfall during the ascending phase has a probability of around 0.773, which is not uncommon. In this season, the ascending phase seems to witness more extreme rainfall occurrences than the descending phase. 
In contrast to the JF and MAM seasons, the odd cycles experienced more excess and deficit years of rainfall than the even ones. 

\begin{table*}[!ht]
   \centering
  \caption{Solar cycles 14-24 classification of phases, during JJAS season}
     \label{phasejjas}
    \begin{tabular}{ccccc} \toprule
     Solar cycle & Minimum & Maximum & Increasing & Decreasing \\ \midrule
       14 & 1901-1902 & 1905-1908 & 1903-1904 & 1909-1912 \\
       15 & 1913-1914 & 1917-1919 & 1915-1916 & 1920-1922 \\
       16 & 1923 & 1928 & 1924-1927 & 1929-1932 \\
       17 & 1933-1934 & 1937-1939 & 1935-1936 & 1940-1943 \\
       18 & 1944 & 1947-1949 & 1945-1946 & 1950-1953 \\
       19 & 1954-1955 & 1956-1959 & 1955 & 1960-1963 \\
    20 & 1964 & 1967-1970 & 1965-1966 & 1971-1975\\
       21 & 1976 & 1979-1981 & 1977-1978 & 1982-1985 \\
      22 & 1986 & 1989-1991 & 1987-1988 & 1992-1995  \\
       23 & 1996 & 1999-2002 & 1997-1998 & 2003-2007   \\
       24 & 2008-2009 & 2011-2014 & 2010 & 2015-2018 \\ \bottomrule  
  \end{tabular}
   \end{table*}

\subsection{OND season}
During the OND season, 11 excess and 15 deficit rainfall events are noticed and are listed below
\begin{itemize}
\item Excess (E) years: 1902, 1930, 1932, 1939, 1944, 1946, 1966, 1977, 1997, 2006, 2010
\item Deficit (D) years: 1904, 1908, 1927, 1938, 1947, 1949, 1967, 1974, 1982, 1985, 1988, 1995, 2000, 2012, 2016 
\end{itemize}   

Figure \ref{extremeond} illustrates the excess and deficient rainfall years during the OND season with different solar phases. Out of 11 excess rainfall years observed, 5 occurred during the descending phase and 6 during the ascending phase of the solar cycles considered. In the case of 15 deficit rainfall years, 6 occurred during the descending phase and 9 during the ascending phases of the solar cycles. The probability of occurrence of 6 years of excess rainfall during the ascending phase is around 0.725 and occurrence of 9 years of deficit rainfall during the ascending phase is around 0.849, which is not unusual. Similar to the JJAS season, the ascending phase and odd cycles record more excess and deficit years of rainfall.

\begin{table*}[!ht]
   \centering
  \caption{Solar cycles 14-24 classification of phases, during OND season}
     \label{phaseond}
    \begin{tabular}{ccccc} \toprule
       Solar cycle & Minimum & Maximum & Increasing & Decreasing \\ \midrule
       14 & 1901-1902 & 1905-1907 & 1903-1904 & 1908-1912 \\
       15 & 1913 & 1917-1918 & 1914-1916 & 1919-1922 \\
       16 & 1923 & 1925-1929 & 1924 & 1930-1932 \\
       17 & 1933-1934 & 1936-1938 & 1935 & 1939-1942 \\
       18 & 1943 & 1946-1949 & 1944-1945 & 1950-1952 \\
       19 & 1953-1954 & 1956-1958 & 1955 & 1959-1963 \\
    20 & 1964 & 1967-1970 & 1965-1966 & 1971-1974\\
       21 & 1975-1976 & 1979-1981 & 1977-1978 & 1982-1984 \\
      22 & 1986 & 1988-1991 & 1987 & 1992-1995  \\
       23 & 1996 & 1999-2002 & 1997-1998 & 2003-2007   \\
       24 & 2008-2009 & 2011-2014 & 2010 & 2015-2018 \\ \bottomrule  
  \end{tabular}
   \end{table*}
   
The possible influence of different phases of solar activity on extreme rainfall events over Kerala was studied. The descending phase was found to play a significant role in the occurrence of excess rainfall in the JF and MAM seasons, and the ascending phase was found to play a significant role in the occurrence of excess rainfall in the JJAS and OND seasons. Due to the magnetic polarity of the sunspots differing between the alternate solar cycles, the study was carried out for the even and old cycles. The even cycles witnessed more extreme rainfall occurrences during the JF and MAM seasons, while the odd cycles during the JJAS and OND seasons. 

Recent studies contribute to the relation of different solar phases with extreme space weather events \citep{Chapman2020,Owens2021}. However, this is the first time the potential relationship between the solar phases and intense rainfall events has been investigated in detail. However, several earlier Indian investigations have documented these rainfall episodes during solar maximum and minimum times.
 \cite{bhalme1981cyclic} observed that the Flood Area Index over India was associated with the double sunspot cycle. 
During alternate solar cycles, \cite{Ananthakrishnan1984} noted significantly more excess rainfall years during the ascending phase. 
According to \cite{jain1997correlation}, the periodicity of floods and droughts are well correlated with sunspot main periods and quasi-periods in the Udaipur subtropical region of Rajasthan. \cite{Bhattacharyya2005} have shown that high rainfall correlates with high solar activity, while low rainfall correlates with low solar activity. Considering the sub-divisions from west central and peninsular India, \cite{Azad2011} reported that the maxima of even sunspot cycles coincided with excess rainfall (with +1 year error), and the minima of odd sunspot cycles coincided with deficit rainfall (with $\pm$2 year error). 

Globally, studies have examined how solar activity impacts extreme rainfall events. \cite{mitchell1979,cook1997} observed that the drought cycle is related to the double (Hale) sunspot cycle in the United States. \cite{Vaquero2004SolarSI} noted that the probability of floods increased during the episodes of high solar activity while studying the floods at the Tagus river basin, Central Spain. A study on Lake Victoria levels in East Africa showed that solar activity affects levels through rainfall. Rainfall maxima lagged one year behind sunspot maxima, resulting in lake level maxima \citep{Stager2007}. Sunspot number revealed a direct correlation with the flood/drought of the Second Songhua river basin, China \citep{hong2015}.  In typical regions of the Loess Plateau, in Yan'an, China, studies of precipitation responses to the solar activity found that maximum precipitation was observed during solar maximum and was associated with solar activity \citep{Li2017}. \cite{YU2019} observed that droughts and floods in the Southern Chinese Loess Plateau were synchronous with solar activities, at least on decadal timescales. 

In some cases, contradictory results were reported.  \cite{Wirth2013} noted that flood frequency in the European Alps increased during cool periods, which coincided with low solar activity. The frequency of flood years is relatively high when solar activity is low, and vice versa, according to \cite{Rimbu2021} studies on River Ammer floods in Germany.  \cite{Li2023} studied the time-lag correlation between solar activity and summer precipitation in the mid-lower reaches of the Yangtze River, China. It was noted that the sunspot number negatively correlates with rainfall, with an 11-month time lag.

Concerns about the solar impact on rainfall patterns across the world have been around for a while, and numerous complex mechanisms that may exist have been examined in various studies \citep{Li2023} and similar periods in the time series of rainfall and solar activity suggested a potential relationship between them \citep{Nitka2019, HEREDIA2019105094}. Variations in Total Solar Irradiance (TSA) \citep{Soon1996} and the solar radiation reaching upper layers of the Earth's atmosphere alters atmospheric circulations causing variations in the rainfall pattern \citep{Baldwin2005,Kodera2007,rycroft}. Solar activity affects the production of cloud condensation nuclei \citep{Svensmark2007,svensmark2019force} and, eventually, affects rainfall distribution. Other natural factors such as El Ni{\~{n}}o/Southern Oscillation (ENSO) and North Atlantic Oscillation (NAO) are some indices that play a crucial role in influencing rainfall and are in turn modulated by solar activity \citep{Leamon2021,kuroda}.

\section{Conclusions} \label{conclusions}
Solar activity affects rainfall differently depending on different regions' climate, topography, and atmospheric conditions. So, the influence of solar activity on Earth's climate has been an interesting research topic for a long. This work examined the possible impact of sunspot numbers on the rainfall over Kerala using 122 years of data. Both annual and seasonal scales were considered by separating the solar and rainfall data into the winter, pre-monsoon, monsoon, and post-monsoon seasons. The climatic effects have been studied by taking 31-year mean values and performing correlation and wavelet analysis (XWT and WTC). The winter and post-monsoon seasons revealed a negative correlation, while pre-monsoon and monsoon seasons showed positive correlations, all with significance. Cross-wavelet transform (XWT) was utilised to study the temporal relationship between sunspot number and rainfall values, and significant cross-power at 8-12 years scale was observed during all the seasons. Wavelet coherence between the two data sets yielded considerable correlation at the 2-4 and 4-8 year scales throughout the four seasons. Strong connections can be seen at longer periods, such as the 8-16 year scale in monsoon and post-monsoon seasons and beyond the 32-year scale in winter and pre-monsoon seasons. The cross-wavelet and coherence analyses did not reveal significant phase information, as consistent phase reversal was observed across all seasons throughout the entire study period.

The association between different phases of solar activity and the extreme precipitation events over Kerala, India, was also evaluated in this study. The solar phases (minimum, increasing, maximum, and decreasing) of Solar Cycles 14-24 were determined for all the seasons. 
The years when Kerala saw excessive or insufficient rainfall were identified, and its relationship with the different phases was investigated.
The descending phase was observed to impact the excess rainfall events (excess/deficit) in the winter and pre-monsoon seasons. In contrast, the ascending phase had a notable effect during the monsoon and post-monsoon seasons, i.e., different phases influence different times of the year. The study focused on the different magnetic polarities of sunspots in alternating solar cycles, explicitly examining the even and odd cycles. Extreme rainfall events were more frequent during the winter and pre-monsoon seasons in the even cycles, whereas in the odd cycles, they were more prevalent during the monsoon and post-monsoon seasons. These findings are presented for the first time and may offer fresh insights into how different phases affect rainfall. This study raises the possibility of a physical link between solar activity and extreme precipitation in Kerala, which aids in predictability and future climate change studies.



\section*{Acknowledgements}
First author acknowledges the financial assistance from the
University Grants Commission (UGC), India, under Savitribai Jyotirao Phule Fellowship for Single Girl Child (SJSGC) (F. No. 82-7/2022(SA-III) dated 07/02/2023). Second author  acknowledges the financial assistance from Department of Science and Technology (DST), Ministry of Science and Technology, India  under INSPIRE Fellowship (Award Letter No. IF180235 dated 08/02/2019).


\bibliographystyle{jasr-model5-names}
\biboptions{authoryear}
\bibliography{refs}

\end{document}